\documentclass[%
showpacs,preprintnumbers,
 amsmath,amssymb,
 aps,
 prd,
 longbibliography,
floatfix,
 lengthcheck,%
]{revtex4}

\usepackage{graphicx}
\usepackage{hyperref}

\begin{document}

\title{Finite size source effects and the correlation of neutrino transition probabilities through supernova turbulence}

\author{James P. Kneller}
\email{jpknelle@ncsu.edu}
\affiliation{Department of Physics, North Carolina State University, Raleigh, North Carolina 27695, USA}

\author{Alex W. Mauney}
\email{awmauney@ncsu.edu}
\affiliation{Department of Physics, North Carolina State University, Raleigh, North Carolina 27695, USA}

\begin{abstract}
The transition probabilities describing the evolution of a neutrino with a given energy along some ray through a turbulent supernova are random variates unique to each ray. 
If the source of the neutrinos were a point then all neutrinos of a given energy and emitted at the same time which were detected in some far off location would have seen the same turbulent profile 
therefore their transition probabilities would be exactly correlated and would not form a representative sample of the underlying parent transition probability distributions. 
But if the source has a finite size then the profiles seen by neutrinos emitted from different points at the source will have seen different turbulence and the correlation of the transition probabilities 
will be reduced. In this paper we study the correlation of the neutrino transition probabilities through turbulent supernova profiles as a function of the separation $\delta x$ between the emission points 
using an isotropic and an anisotropic power spectrum for the random field used to model the turbulence.
We find that if we use an isotropic power spectrum for the random field, the correlation of the high (H) density resonance mixing channel transition probability is significant, greater than 0.5, for emission separations of $\delta x=10\;{\rm km}$, typical of proto neutron star radii, only when the turbulence amplitude is less than $C_{\star} \sim 10\%$; at larger amplitudes the correlation in this channel drops close to zero for this same separation of $\delta x=10\;{\rm km}$. 
In contrast, there is significant correlation in the low (L) density resonant and non-resonant channels even for turbulence amplitudes as high as 50\%.     
Switching to anisotropic spectra requires the introduction of an `isotropy' parameter $k_{I}$ whose inverse defines the scale below which the field is isotropic. We find 
the correlation of all transition probabilities, especially the H resonance channel, strongly depends upon the choice of $k_{I}$ relative to the long wavelength radial cutoff $k_{\star}$. 
The spectral features in the H resonance mixing channel of the next Galactic supernova neutrino burst may be strongly obscured by large amplitude turbulence when it enters the signal due to the finite 
size of the source while the presence of features in the L and non resonant mixing channels may persist, the exact amount depending upon the degree of anisotropy of the turbulence. 
\end{abstract}

\pacs{47.27.-i,14.60.Pq,97.60.Bw}
\date{\today}

\maketitle


\section{Introduction}

The neutrino signal from the next core-collapse supernova in our Galaxy will give us an unprecedented opportunity to peer into the heart of an exploding star and confront our 
current paradigm of how these stars explode with observations. But decoding the message will be no easy task because the neutrino signal will have experienced so many flavor-changing events on the trip from proto-neutron star to our detectors that scramble the information, see for example Kneller, McLaughlin \& Brockman \cite{Kneller:2007kg} and Lund \& Kneller \cite{LK2013}. The first flavor changing effect the signal experiences is due to neutrino self interactions / collective effects in the region up to $\sim 1000\;{\rm km}$ above the proto neutron star \cite{Pantaleone:Gamma1292eq,Samuel:1993uw,Pastor:2001iu,Duan:2005cp,Hannestad:2006nj,Duan:2006an,Raffelt:2007cb,2008PhRvD..78l5015R,Duan:2009cd,Duan:2010bg,2011PhRvL.106i1101D,2011PhRvL.107o1101C,2011PhRvD..84h5023R,2012JPhG...39c5201G,2012PhRvL.108z1104C,2012PhRvD..85k3007S,2012PhRvL.108w1102M} followed by the Mikheyev, Smirnov \& Wolfenstein (MSW) \cite{M&S1986,Wolfenstein1977} effect which is complicated in supernovae by the impression of the shockwave 
racing through the stellar mantle \cite{Schirato:2002tg,Takahashi:2002yj,Fogli:2003dw,Tomas:2004gr,Choubey:2006aq,Kneller:2007kg,Gava:2009pj,LK2013}. Turbulence in the mantle, seeded during the earlier neutrino heating/Standing Accretion Shock Instability phase \cite{2011ApJ...742...74M,2012arXiv1210.5241D,2012arXiv1210.6674O,2012ApJ...755..138H,2012ApJ...746..106P,2012ApJ...761...72M,2012ApJ...749...98T,2013arXiv1301.1326L}, also needs to be included usually by modelling \cite{Loreti:1995ae,Fogli:2006xy,Friedland:2006ta,2010PhRvD..82l3004K,2011PhRvD..84h5023R}. Finally, there is the possibility of Earth matter effects leaving an imprint in the signal though a recent study expects this effect to be minimal \cite{2012PhRvD..86h3004B}.  

What makes decoding the signal even more of a challenge is that the neutrinos we receive at a given instant and with a given energy will not have experienced the same flavor evolutionary history. 
The neutrinos arriving at a detector will have been emitted from different locations at the source and both the neutrino collective and the MSW+turbulence effects will vary from trajectory to trajectory. 
Starting with Duan \emph{et al.} \cite{Duan:2005cp}, the self interaction effects in calculations where the neutrino emission over the source is assumed to be spherically symmetric have been seen to be `angle dependent' in the sense that a neutrino following a pure radial trajectory differs from one emitted at an angle relative to the normal. Presumably allowing for aspherical source emission would only make the trajectory dependence even stronger. 
Similarly the MSW plus turbulence effects are also trajectory dependent. If we temporarily cast aside the turbulence and focus on the gross structure of the explosion i.e.~ the lowest angular multipole moments, an aspherical passage of the shock through the star, by itself, leads to a line-of-sight dependence. But, one must recall that we will not observe the neutrinos from a supernova at widely different lines of sight, all our detectors are here on Earth. The size of the source is of order the proto-neutron star radius, i.e.~ $\sim 10\;{\rm km}$ while the shock effects show up in the signal when the shock has propagated out to $r \sim 10^{4}\;{\rm km}$. As long as the curvature of the shock is over a lengthscale greater than the source size the neutrinos which appear in our detectors will all have seen essentially the same profile. When one re-inserts the turbulence into the profile, one realizes this approximation may no longer be valid because turbulence extends to much smaller lengthscales even when the shock is far out in the stellar mantle. The density profile along two, essentially parallel lines of sight to a distant detector separated by $\sim 10\;{\rm km}$ will no longer be negligibly dissimilar and one must consider how the dissimilarity of the profiles propagates to the neutrinos. Any correlation will lead to a potential new feature of the neutrino signal. 

It has been shown that the transition probabilities for a single neutrino - the set of probabilities that relates
the initial state to the state after passing through the supernova - is not unique when turbulence is inserted into a profile: it will depend upon the exact turbulence pattern seen 
by the neutrino as it travelled through the supernova \cite{Friedland:2006ta,2010PhRvD..82l3004K,2013arXiv1302.3825K}. Those transition probabilities are drawn from distributions whose properties will 
depend upon the stage of the explosion, the character of the turbulence, and the neutrino energy and mixing parameters. If the coherence of two neutrinos emitted at the same time and with the same energy but from different
locations is small then the final states are uncorrelated and one would expect that the flux at a detector would just be the mean of whatever distribution
describes the transition probabilities multiplied by the initial spectra. But if the coherence is high then all the neutrinos will have the same set of transition probabilities which one might expect to 
`scintillate' together as the turbulence evolves. Of course, this ignores the issue of energy resolution and temporal binning of the signal that becomes necessary because of the limited statistics.  

The purpose of this paper is to consider the issue of finite source size and the correlation of the neutrino transition probabilities along parallel trajectories through turbulent supernova profiles. Our calculations expand upon the work of Kneller \& Volpe \cite{2010PhRvD..82l3004K} and Kneller \& Mauney \cite{2013arXiv1302.3825K} upon which we rely heavily for the techniques used to calculate the turbulence effects and as context for our results.
We begin by describing the calculations we undertook paying particular attention to the construction of the random fields used to model the turbulence.  
The basic approach to determining the effects of turbulence are then demonstrated, followed by the computation of the transition probability correlation as a function of the 
separation between the emission points. We finish by summarizing our findings and discuss the implications for the Galactic neutrino burst signal.


\section{Description of the calculations}

The neutrino transition probabilities are the set of probabilities of measuring some neutrino state $\nu_{i}$ given an initial neutrino state $\nu_{j}$ i.e~ $P(\nu_{j} \rightarrow \nu_{i}) = P_{ij}$. We shall denote 
antineutrino transition probabilities by $\bar{P}_{ij}$. 
If the $S$-matrix relating the initial and final wavefunctions is known then these probabilities are just the square amplitudes of the elements of $S$. 
The $S$-matrix is calculated from the Schrodinger equation
\begin{equation}
\imath \frac{dS}{dr} = H\,S
\end{equation} 
where $H$ is the Hamiltonian. The Hamiltonian is the sum of the vacuum contribution $H_0$ and the MSW potential $V$ which describes the 
effect of matter. The vacuum Hamiltonian is diagonal in what is known as the `mass' basis and in this basis $H_{0}$ is defined by two mass squared differences $\delta m_{ij}^2 = m_i^2 - m_j^2$ and the neutrino energy
$E$.
The mass basis is related to the flavor basis by the Maki-Nakagawa-Sakata-Pontecorvo \cite{Maki:1962mu,Nakamura:2010zzi} unitary matrix $U$. The most common parametrization of $U$ is in terms of 
three mixing angles, $\theta_{12}$, $\theta_{13}$ and $\theta_{23}$, a CP phase and two Majoranna phases.    

The MSW potential $V$ is diagonal in the flavor basis because matter interacts with neutrinos based on their flavor. The neutral current interaction leads to a contribution to $V$ which is 
common to all flavors. This may be omitted because it leads only to a global phase which is unobservable. The charged current potential only affects the electron flavor neutrino and antineutrinos and 
is given by $\sqrt{2} G_F n_e({\bf r})$ where $G_F$ is the Fermi constant and $n_e({\bf r})$ the electron density. 

In matter the two contributions to $H$ means neither the mass nor the flavor states diagonalize the matrix. But there is a basis known as the matter basis which does diagonalize $H$ i.e.\ for a given value of the electron density there is a matrix $\tilde{U}$ such that $\tilde{U}^{\dagger} H \tilde{U} = K$ where $K$ is the diagonal matrix of eigenvalues. When the MSW potential vanishes the matter basis  becomes the mass basis up to arbitrary phases. 
The matter basis is the most useful for studying the evolution of neutrinos through matter because it removes the trivial adiabatic MSW transition and it will be 
the basis we use to report our results in this paper. 
We refer the reader to Kneller \& McLaughlin \cite{2009PhRvD..80e3002K} and Galais, Kneller \& Volpe \cite{2012JPhG...39c5201G} for a more detailed description 
of the matter basis.  

\begin{figure}
\includegraphics[clip,width=\linewidth]{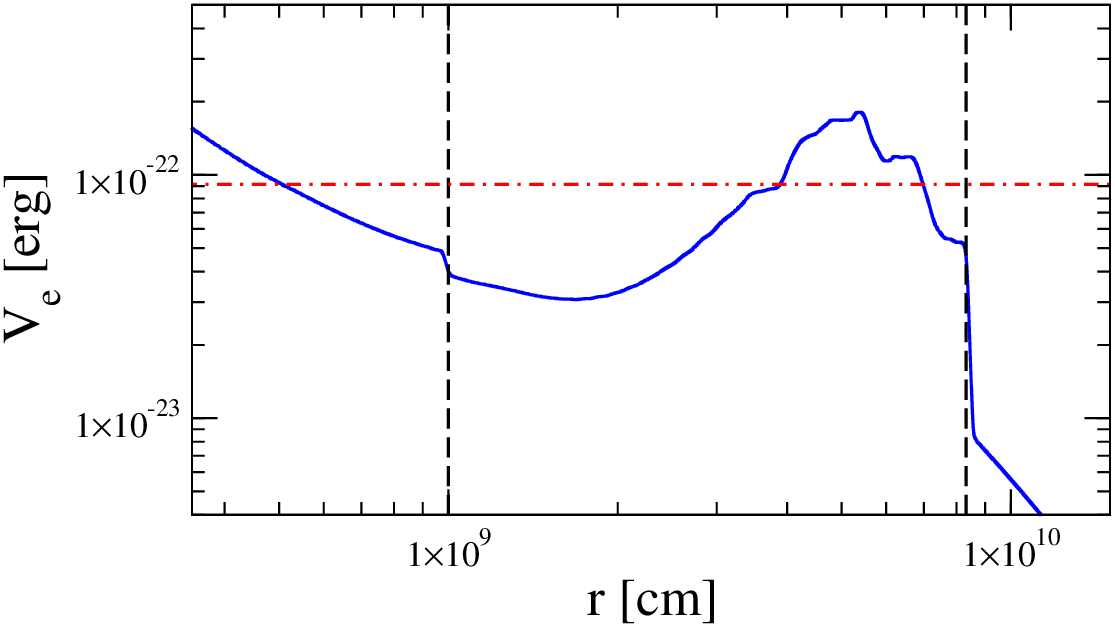}
\caption{The turbulence free MSW potential as a function of distance through a supernova taken from a hydrodynamical simulation. The vertical lines indicate the positions of
the reverse and froward shock in the profile. The horizontal dashed-dotted line is the two-flavor resonance density for a $25\;{\rm MeV}$ neutrino with mixing angle
$\sin^2 2\theta = 0.1$ and mass splitting $\delta m^{2} = 3\times 10^{-3} \;{\rm eV^2}$ \label{fig:profile}}
\end{figure}

We now turn our attention to the turbulent density profiles through which we shall send our neutrinos.
As usual, we shall model the turbulence by multiplying a turbulence free density profile by a Gaussian random field.  
Since the spatial extent of the neutrino emission, of order $10\;{\rm km}$, is much smaller than 
the radial location of the turbulence, of order $r\sim 10^{4}-10^{5}\;{\rm km}$, we shall ignore any curvature of the density profile features and
use a plane-parallel model for the supernova. The z axis of our Cartesian co-ordinate system is aligned with the radial direction of the profile. 
The profile we adopt is from a one-dimensional hydrodynamical simulation of a supernova taken from Kneller, McLaughlin \& Brockman \cite{Kneller:2007kg}. 
This profile is shown in figure (\ref{fig:profile}) and is the same one used in Kneller \& Mauney \cite{2013arXiv1302.3825K}. 
The figure shows the presence of two shocks: the forward shock at $r_s$ and the reverse shock at $r_r$. 
In multi-dimensional simulations of supernova both these shock fronts are aspherical and fluid flow through the distorted shocks leads to strong turbulence in the region between them. 
Our selection of this profile also determines the neutrino energy we shall use since we wish the neutrinos to have an H resonance density that does not intersect the shocks. 
Therefore we pick $25\;{\rm MeV}$ for the neutrino energy and the reader may observe that the two-flavor resonance density for a $25\;{\rm MeV}$, shown in the figure, does not 
intersect the shocks as required.   

The turbulence is inserted by multiplying the profile in the region between the reverse and forward shocks by a factor $1+ F({\bf r})$ where $F({\bf r})$ is a three-dimensional Gaussian random field with zero mean. The random field is represented by a Fourier series, that is 
\begin{equation}\label{eq:F3D}\begin{split}
F({\bf r})&=C_{\star}\,\tanh\left(\frac{r-r_r}{\lambda}\right)\,\tanh\left(\frac{r_s-r}{\lambda}\right) \\
& \times\sum_{n=1}^{N_k}\,\sqrt{V_{n}}\left\{ A_{n} \cos\left({\bf k}_{n}\cdot {\bf r}\right) + B_{n} \sin\left({\bf k}_{n}\cdot {\bf r}\right) \right\}. 
\end{split}
\end{equation}
In this equation the parameter $C_{\star}$ sets the amplitude of the fluctuations while the two $\tanh$ functions 
are included to suppress fluctuations close to the shocks and prevent discontinuities. The  parameter $\lambda$ is a damping scale which we set to $\lambda=100\;{\rm km}$. 
The random part of $F$ appears in the the second half of equation (\ref{eq:F3D}) because the set of co-efficients $\{A\}$ and $\{B\}$ are independent standard Gaussian random variates with zero mean. 
The ${\bf k}_{n}$ are a set of wavenumbers and, finally, the paramaters $V_{n}$ are k-space volume co-efficients. 
The method of fixing the $N_k$ $k$'s, $V$'s, $A$'s and $B$'s for a realization of $F$ is `variant C' of the Randomization Method
described in Kramer, Kurbanmuradov, \& Sabelfeld \cite{2007JCoPh.226..897K} which we have generalized to three dimensions.
The Randomization Method in general partitions the space of wavenumbers into $N_{k}$ regions and from each we select a random wavevector using the power-spectrum, $E({\bf k})$, as a
probability distribution. The volume paramaters $V_{n}$ are the integrals of the power spectrum over each partition if the power spectrum is normalized to unity.
Variant C of the Randomization Method divides the k-space so that the number of partitions per decade is uniform over $N_d$ decades starting from a cutoff scale $k_{\star}$.
Throughtout this paper we shall use a wavenumber cutoff $k_{\star}$ set to twice the distance between the
shocks i.e.~ $k_{\star}=\pi/(r_s-r_r)$. The logarithmic distribution of the modes increases the efficiency of the algorithm in the sense that we can use a `small' value of $N_k$ and also 
the agreement between the exact statistical behavior of the field and
that of an ensemble of realizations is uniform over some range of lengthscales i.e.~ it is scale invariant. This feature is
important for our study because the oscillation wavelength of the neutrinos is constantly changing as the density evolves. 
The minimum lengthscale we need to cover has been shown by Friedland \& Gruzinov \cite{Friedland:2006ta} and Kneller, McLaughlin \& Patton \cite{2012arXiv1202.0776K} 
to be the reduced oscillation wavelengths for the neutrinos and antineutrinos i.e.~ $\lambda_{ij} =1/|\delta k_{ij}|$ and $\bar{\lambda}_{ij}
=1/|\delta \bar{k}_{ij}|$ - where $\delta k_{ij}$ and $\delta\bar{k}_{ij}$ are the differences between the eigenvalues $i$ and $j$ of the neutrinos and
antineutrinos respectively. Kneller \& Mauney \cite{2013arXiv1302.3825K} showed the wavelengths in the turbulence region were of order $1\;{\rm km}$ or greater which is approximately 
four orders of magnitude smaller than the shock separation. This means we need to pick $N_d \geq 4$ to cover the necessary decades in k-space.

\subsection{The power spectrum}

The final component of our calculations we have yet to discuss is the power spectrum $E({\bf k})$. 
In this paper we shall consider two power spectra and our first choice, due to its simplicity, is a normalized three-dimensional, isotropic inverse power-law spectrum given by 
\begin{equation}
E({\bf k}) = \frac{(\alpha-1)}{4\pi k_{\star}^{3}}\,\left(\frac{k_{\star}}{|k|}\right)^{\alpha+2}\,\Theta(|k|-k_{\star}). \label{eq:iostropicps}
\end{equation}
for $|k| \geq k_{\star}$ where $|k|$ is the magnitude of the wavevector ${\bf k}$. Throughout this paper we shall adopt the Kolmogorov spectrum where
$\alpha=5/3$. The one dimensional power spectrum for the $k_z$ component of the wavevector is 
\begin{eqnarray}
E_{1}(k_z) & = & \frac{(\alpha-1)}{2\alpha k_{\star}}\,\left(\frac{k_{\star}}{|k_z|}\right)^{\alpha}\,\Theta(|k_z|-k_{\star}) \nonumber \\ 
& & + \frac{(\alpha-1)}{2\alpha k_{\star}}\,\Theta(k_{\star}-|k_z|) \label{eq:1Diostropicps}
\end{eqnarray}
which differs from the one dimensional power spectrum used by Kneller \& Mauney \cite{2013arXiv1302.3825K} because for $|k_z|\geq k_{\star}$ the power is
suppressed by the factor $1/\alpha$ and the one dimensional spectrum is non-zero for $|k_z| \leq k_{\star}$. 
The two-point correlation function $B(\delta {\bf r})$ for this choice of a power spectrum depends only the magnitude of the separation, $\delta r$, and may be calculated analytically to be 
\begin{widetext}
\begin{equation}
B(\delta r) = \frac{\imath \,(\alpha-1)}{2}\,\left(2\pi\,k_{\star}\,\delta r \right)^{\alpha-1} \left\{\exp\left(\frac{\imath\pi\alpha}{2}\right)\,\Gamma(-\alpha,2\imath\pi\,k_{\star} \,\delta r) -
\exp\left(\frac{\imath\pi\alpha}{2}\right)\,\Gamma(-\alpha,-2\imath\pi\,k_{\star}\,\delta r) \right\}
\end{equation}
\end{widetext}
where $\Gamma(n,x)$ is the incomplete Gamma function. 
There is one last quantity to determine: the number of $N_k$ of elements in the sets of random wavenumbers, coefficients and volumes. 
To find this quantity we compare the statistical properties of an ensemble of random field realizations with the exact expressions as a function of 
the ratio $N_k/N_d$ for a given $N_d$. The statistical property we compute is the second order structure function $G_2(\delta {\bf r} )$ which is given by 
\begin{equation}
G_2(\delta {\bf r}) = \langle F({\bf r} + \delta {\bf r}) - F({\bf r}) \rangle^2 \label{eq:G2}
\end{equation}
where $\delta {\bf r}$ is the separation between two points. The function $G_2(\delta{\bf r})$ is related to the two-point correlation function $B(\delta
{\bf r})$ via $G_2(\delta {\bf r})/2 = 1-B(\delta {\bf r})$. For the isotropic power spectrum both $G_2$ and $B$ are only functions of the magnitude of $\delta {\bf r}$ 
and the correlation function is given above.  
\begin{figure}[b]
\includegraphics[clip,width=\linewidth]{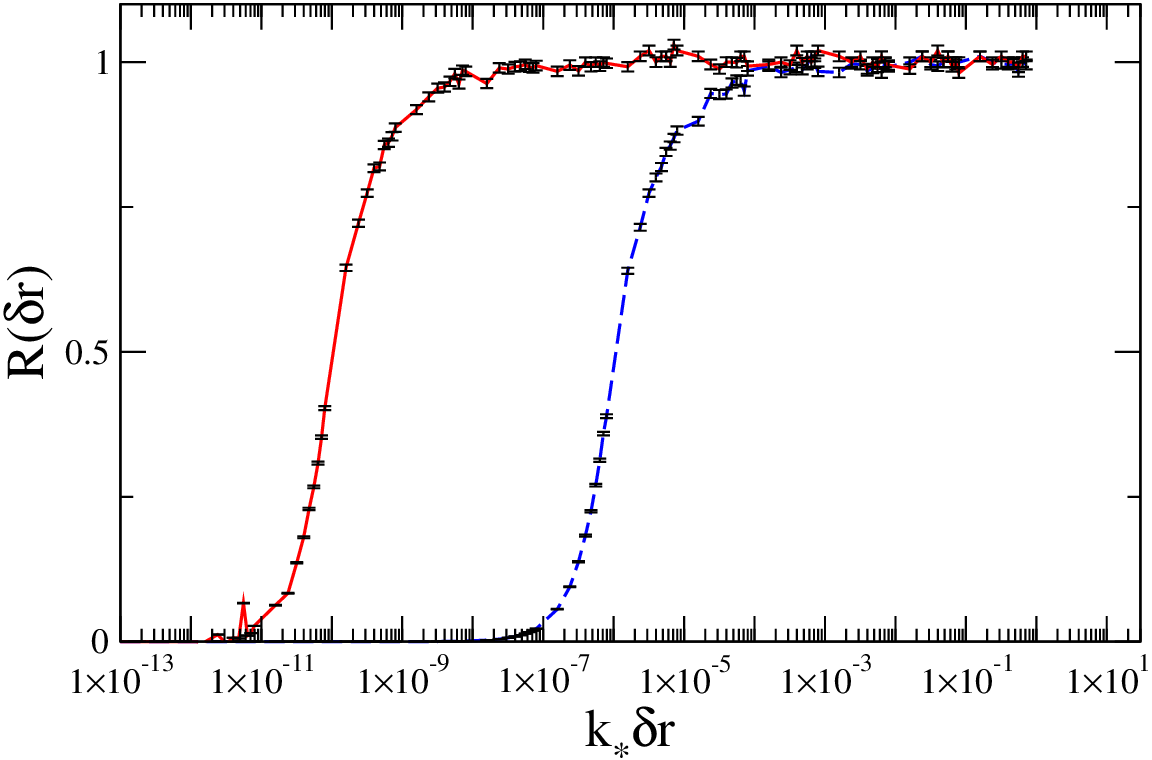}
\caption{The ratio of the structure function $G_2(\delta r)$ as a function of $k_{\star} \delta r$ for two randomly orientated points in a 3-D homogeneous and
isotropic Gaussian random field to the exact structure function. The two curves in the figure correspond to $\{N_k,N_d\}=\{50,5\}$ (blue solid) and $\{N_k,N_d\}=\{90,9\}$.
At every $k_{\star} \delta r$ we generated $30,000$ realization of the field and the error bar on each point is the standard deviation of the mean $F({\bf r} + \delta {\bf r}) - F({\bf r})$.. \label{fig:GRF3Diso}} 
\end{figure}
In figure (\ref{fig:GRF3Diso}) we show the ratio $R(\delta r)$ of the numerically calculated structure function for the isotropic random field to the exact solution as a function of
the scale $k_{\star}\delta r$ when we use either $N_{k}=50$ wavenumbers spread over $N_d=5$ decades or $N_k=90$ wavenumbers over $N_d=9$ decades. The numerical
calculation is the average of $30,000$ realizations of the turbulence and the error bar on each point is the error on the sample mean. The figure indicates
that the method we use to generate random field realizations reproduces the analytic results for the structure function very well and with high efficiency
because good agreement between the statistics of the ensemble and the exact result requires just $N_{k}/N_d = 10$. In fact we find even $N_{k}/N_d$ ratios of 
just $N_{k}/N_d \sim 2-3$ are sufficient to give acceptable agreement but we re-assure the reader we shall stick with $N_{k}/N_d = 10$.

But isotropic and homogeneous three-dimensional turbulence is perhaps not a realistic scenario for supernova because the 
gravitational potential and the general fluid flow are in the radial direction. 
Only on sufficiently small scales should the turbulence become isotropic. This division into large and small lengthscales 
indicates we should partion the power-spectrum so that for $|k_z| \geq k_I$ the spectrum is isotropic, where $k_I$ is the isotropy scale, between 
$k_\star \leq |k_z| \leq k_I$ the spectrum is anisotropic and then below the cutoff scale, $|k_z| \leq k_{\star}$, the power spectrum should be set to zero
since there should be no modes on scales larger than $1/k_{\star}$. For $|k_z| \geq k_{I}$ where the spectrum is isotropic we use a power spectrum resembling equation (\ref{eq:iostropicps}) 
\begin{equation}
E({\bf k}) = \frac{\alpha\,(\alpha-1)}{4\pi k_{\star}^{3}}\,\left(\frac{k_{\star}}{|k|}\right)^{\alpha+2}\;\Theta(|k|-k_{I}). \label{eq:psB} 
\end{equation}
Note the additional factor of $\alpha$ in the numerator. For $k_{\star} \leq |k_z| \leq k_{I}$ we write the spectrum as the product $E(k_{x},k_{y},k_{z})= E(k_{x},k_{y}) \times E(k_{z})$. The spectrum $E(k_{z})$
is chosen to be a continuation of the inverse power-law given above while the spectrum in the xy directions, $E(k_{x},k_{y})$, is the spectrum 
of the isotropic/homogeneous region in these directions fixed at $|k_z|=k_{I}$. The spectrum for $k_{\star} \leq |k_z| \leq k_{I}$ is thus
\begin{eqnarray}
E(k_{x},k_{y},k_{z}) & = & \frac{\alpha(\alpha-1)}{4\pi k_{\star}^{3}}\left(\frac{k_{\star}}{|k_{z}|}\right)^{\alpha}\left(\frac{k_I^{2}}{k_{x}^{2}+k_{y}^{2}+k_{I}^{2}}\right)^{\alpha/2+1} \nonumber \\ 
& & \;\Theta(k_{I}-|k|)\,\Theta(|k|-k_{\star}). \label{eq:psA}
\end{eqnarray}
The reader may verify the power spectrum defined by equations (\ref{eq:psB}) and (\ref{eq:psA}) is normalized. This anisotropic three-dimensional power spectrum yields a one-dimensional
spectrum along the z direction given by
\begin{equation}
E_1(k_z) = \frac{(\alpha-1)}{2\,k_{\star}} \left( \frac{k_{\star}}{|k_z|}\right)^{\alpha}\,\Theta(|k_{z}|-k_{\star}) \label{eq:1Daniostropicps}
\end{equation}
for $|k_z| \geq k_{\star}$ which is exactly the same as the one-dimensional spectrum used in Kneller \& Mauney \cite{2013arXiv1302.3825K}. 
There is no analytic formula for the two-point structure function for \emph{randomly} orientated separations using this power spectrum but if we consider
the two-point structure function of the random field for points orientated along the z direction then we can compute that in this direction 
\begin{widetext}
\begin{equation}
B(\delta z) = \frac{(\alpha-1)}{2}\,\left(2\pi\,k_{\star}\,\delta z \right)^{\alpha-1} \left\{\exp\left(\frac{\imath\pi\alpha}{2}\right)\,\Gamma(1-\alpha,2\imath\pi\,k_{\star}\,\delta z) +
\exp\left(\frac{\imath\pi\alpha}{2}\right)\,\Gamma(1-\alpha,-2\imath\pi\,k_{\star}\,\delta z) \right\}. \label{eq:zcorrelation}
\end{equation}
\end{widetext}  
Compared to the isotropic spectrum above, this anisotropic spectrum differs in important ways. 
First, even if we set $k_I=k_{\star}$ we observe that the lack of power in the region $|k_z| \leq k_{\star}$ means we have to compensate 
by increasing the structure / decreasing the correlation by the factor $\alpha$. This increase is the reason for the appearance of the extra factor $\alpha$ in equation (\ref{eq:psB}). 
Next, as we increase the ratio $f_I = k_I/k_{\star}$, we push more and more of the structure of the field in the xy direction to ever smaller scales reducing even further 
the correlation of the field at some fixed non-radial separation $\delta x$ compared to the isotropic case. 
\begin{figure}[b]
\includegraphics[clip,width=\linewidth]{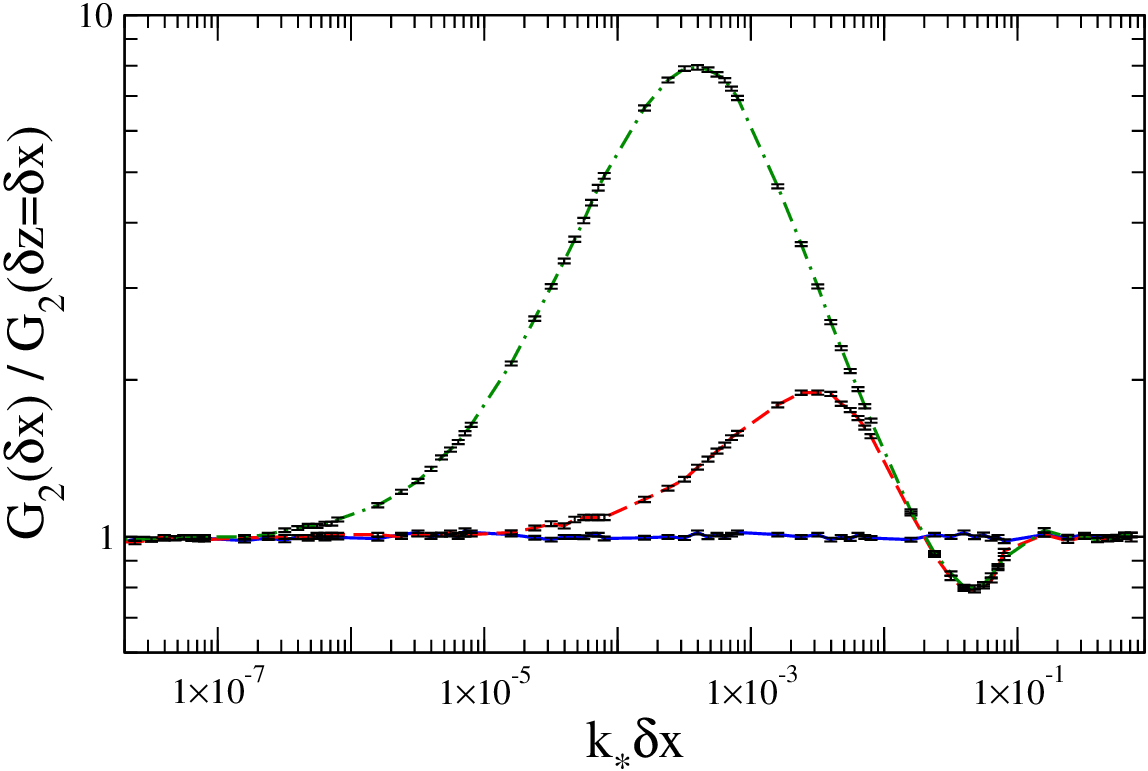}
\caption{The ratio of the structure function $G_2(\delta x)$ as a function of $k_{\star} \delta x$ for two points aligned along the $x$ direction to the 
structure function $G_2(\delta z)$ of two points aligned along the $z$ direction at $\delta x =\delta z$. 
The three curves in the figure correspond to $k_I = k_{\star}$ (solid), $k_I = 10\,k_{\star}$ (dashed) and  $k_I = 100\,k_{\star}$ (dot-dashed).
At every $k_{\star} \delta x$ we generated $30,000$ realization of the field and the error bar on each point is the standard deviation of $F(x + \delta x) -
F(x)$. The structure function $G_2(\delta z)$ was computed using the correlation function given in (\ref{eq:zcorrelation}) and the relationship $G_2(\delta z)/2
= 1-B(\delta z)$. The inputs to the random field generator were $N_k=90$, $N_d=9$.  \label{fig:GRF3Daniso}}
\end{figure}
This extra power at small scales can be seen in figure (\ref{fig:GRF3Daniso}) which is a plot of the ratio of the one-dimensional two-point structure function in the x direction relative to the
structure function along the z direction at the same separation scale for three values of $f_I$. As promised, when  $f_I=1$ there is an equal amount of structure in the field along 
both radial and non-radial directions but as $f_I$ increases we push more and more of the structure of the field in the xy direction to smaller scales. 

The anisotropic power spectrum we have constructed means the turbulence along different parallel rays is less correlated than the turbulence along two rays at the same separation when 
the power spectrum is isotropic. If that's the case then the transition probabilities for the neutrinos travelling along those two rays should also be less correlated and below 
we quantify the decrease.


\section{Results}

Now that we have the random fields to model the turbulence we are all set to generate turbulent profiles and send neutrinos and antineutrinos through them.
To achieve higher efficiency we follow six neutrinos and six antineutrinos simultaneously through every realization of the turbulence with one neutrino and one antineutrino  
emitted at $x \in \{0,10^4,10^5,10^6,10^7,10^8\}\;{\rm cm}$. Each time we generate a new realization we end up with a different set of transition probabilities so by repeating the calculation many times - in our case a minimum of one thousand times but often much larger - we can create an ensemble of transition probabilities of size $N$ from each emission point. Once we have our ensemble we can then go ahead and compute means $\langle P_{ij}(x)\rangle$, variances $V_{ij}(x)$, and, of course, correlations 
\begin{equation}
\rho_{ij}(\delta x) = \frac{\langle P_{ij}(x)P_{ij}(x+\delta x)\rangle - \langle P_{ij}(x)\rangle\,\langle P_{ij}(x+\delta x)\rangle}{\sqrt{V_{ij}(x)\,V_{ij}(x+\delta x)}} \label{eq:rho}
\end{equation}
The correlation of the antineutrino transition probabilities will be denoted as $\bar{\rho}_{ij}$.  
In the large N limit the error on the correlation is expected to be $\sigma_{\rho} = (1-\rho^{2})/\sqrt{N-1}$. 
Combining the results from the six emission points we can form fifteen separations $\delta x$ so fifteen correlations but two points must be remembered: first, groups of them will cluster e.g.\ we will have a value for the correlation at $\delta x =10\;{\rm km}$ but also two more at $\delta x =9\;{\rm km}$ and $\delta x =9.9\;{\rm km}$ and second, these groups of transition probability correlations are themselves correlated - half the data in each correlation value is the same for all members of the cluster. Nevertheless these clusters are useful because they serve as a consistency check - we should expect the results to be similar within each cluster - and also they give us an indication if the error in the results are comparable to the expected, large-N error $\sigma_{\rho}$ given above. 

We also need to specify the neutrino mixing parameters we have used. The hierarchy will be set to normal and we shall comment on how our results translate to the inverted hierarchy. As discussed, the neutrino energy will be fixed at $E=25\;{\rm MeV}$, typical of supernova neutrino energies and we shall set the neutrino mixing parameters to be $\delta m^2_{12}= 8 \times 10^{-5}$eV$^2$, $\delta m^2_{23}= 3 \times 10^{-3}$eV$^2$, $\sin^{2}
2\theta_{12}=0.83$, and $\sin^{2} 2\theta_{23}=1$. The recent measurements of the last mixing angle $\theta_{13}$ by T2K \cite{2011PhRvL.107d1801A}, Double Chooz \cite{2012PhRvL.108m1801A}, RENO \cite{2012PhRvL.108s1802A} and Daya Bay \cite{2012PhRvL.108q1803A} are all in the region of $\theta_{13} \approx 9^{\circ}$. We shall adopt this value for the majority of this paper but this result is sufficiently new that we 
shall show on occasion results with multiple values of $\theta_{13}$ in order to put this result in context.

Finally, the turbulence amplitude $C_{\star}$ will be allowed to vary but we shall focus upon larger values. With the measurement of a large value of $\theta_{13}$ the turbulence effects are negligible for amplitudes of order 
$C_{\star} \sim 1\%$ \cite{2013arXiv1302.3825K}. 


\begin{figure}
\includegraphics[clip,width=\linewidth]{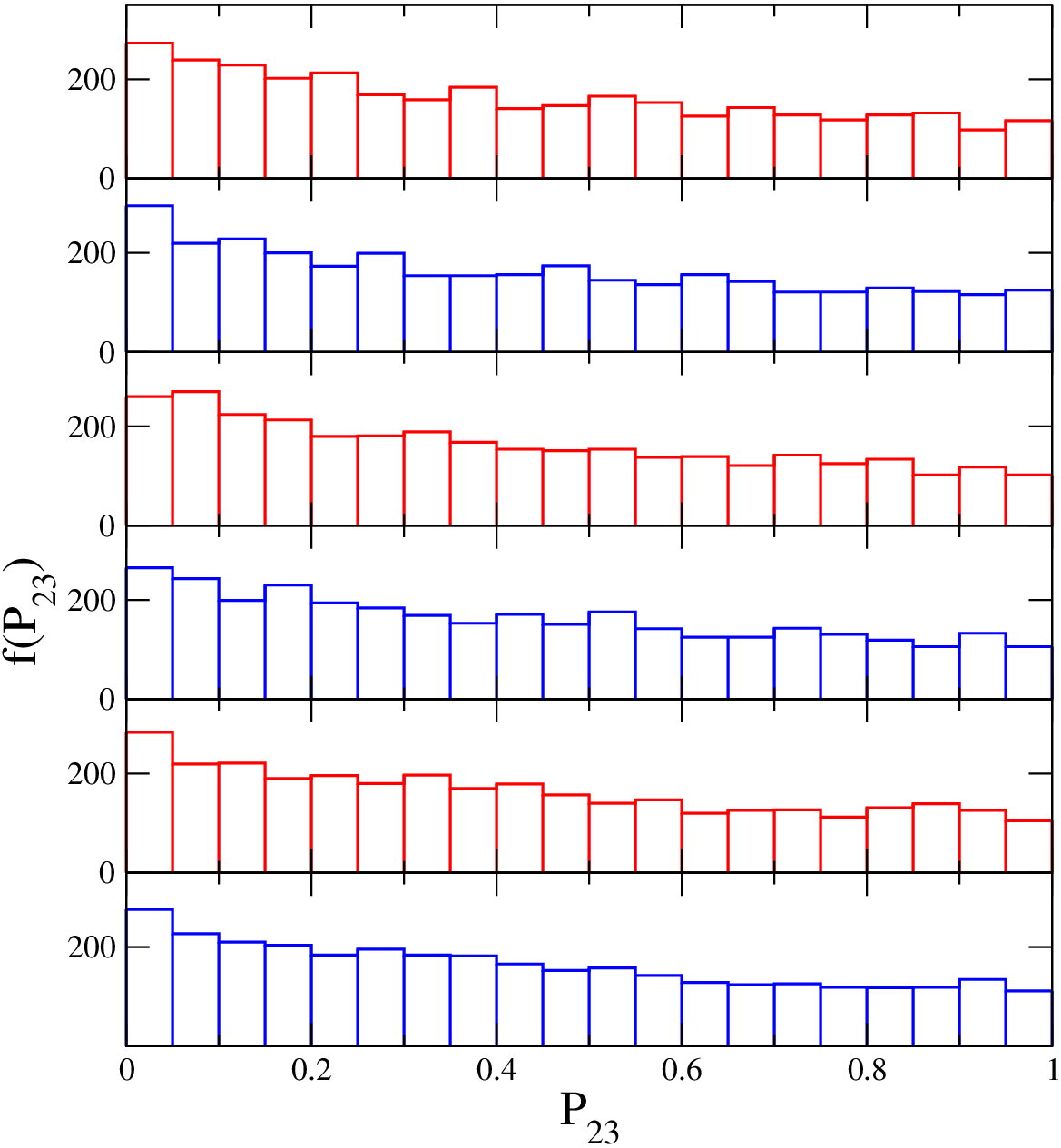}
\caption{The frequency distribution of the transition probability $P_{23}$ for each of the neutrino emission points $x$. From bottom to top the emission points
are $x = 0,\;10^4,\;10^5,\;10^6,\;10^7,\;10^8\;{\rm cm}$. The turbulence amplitude is
set to $C_{\star}=30\%$, we used $N_{k} = 50$, $N_d=5$ for the 3-D turbulence field generator, and the neutrino mixing parameters used are those given in the text with $\sin^2 2\theta_{13}=0.1$. \label{fig:f(P23)} }
\end{figure}
\begin{figure}
\includegraphics[clip,width=\linewidth]{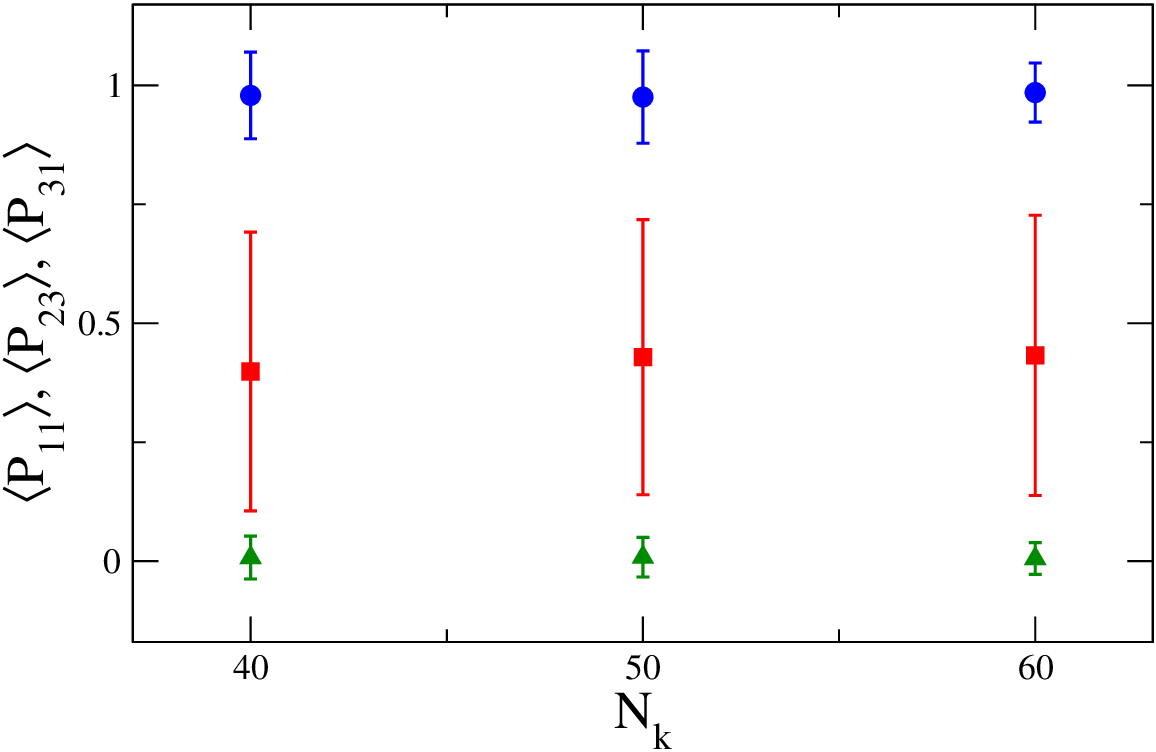}
\caption{The mean of the transition probability $P_{11}$ (circles), $P_{23}$ (squares) and $P_{31}$ (triangles) of the neutrinos emitted at $x=0$ as a function of the parameter $N_{k}$ keeping the ratio $N_{k}/N_{d}$ fixed 
at $N_{k}/N_{d} =10$. The error bars are not the error on the mean but rather the standard deviation of the samples. The turbulence amplitude is set to $C_{\star}=30\%$ and $\sin^2 2\theta_{13}=0.1$\label{fig:meanP23vsNk} }
\end{figure}
\subsection{The point source statistics}
Before we show our results for the correlation of the transition probabilities as a function of the emission separation, we consider first the statistical properties of the ensembles  
for each emission point. In addition to being interesting in their own right and useful as a reference, these calculations allow us to test that our 3D random field generator is working properly because the ensembles for each point of emission should be consistent and independent of $x$. 

In figure (\ref{fig:f(P23)}) we show the frequency distribution of $P_{23}$ for the six emission locations $x$ using the mixing paramaters given above, $C_{\star} = 30\%$, $N_{k} = 50$, $N_d=5$ and $\sin^2(2\theta_{13})=0.1$ and 
the isotropic power spectrum. The sample size is $N=3265$ for each emission point. In each panel of the figure the reader will observe that the transition probability is almost uniformly distributed - there is a slight decrease in the frequency of higher values of $P_{23}$ - but, more importantly, there is no observed trend with $x$. A closer inspection of figure (\ref{fig:f(P23)}) also hints at some correlation: the bottom few panels of the
figure are very similar. We have reproduced this calculation for other choices of the $N_{k}$ and $N_d$ paramaters. The results are shown in figure (\ref{fig:meanP23vsNk}) where we plot the mean values and standard deviations of $P_{11}$, $P_{23}$ and $P_{31}$ for ensembles of neutrinos emitted at $x=0$ as a function of the parameter $N_{k}$ keeping the ratio $N_k/N_d$ fixed at $N_k/N_d=10$. There is no discernable trend with $N_k$ and we march on  confident that setting $N_{k} = 50$ and $N_d=5$ does not bias our results.  
 
\begin{figure*}
\includegraphics[clip,width=0.45\linewidth]{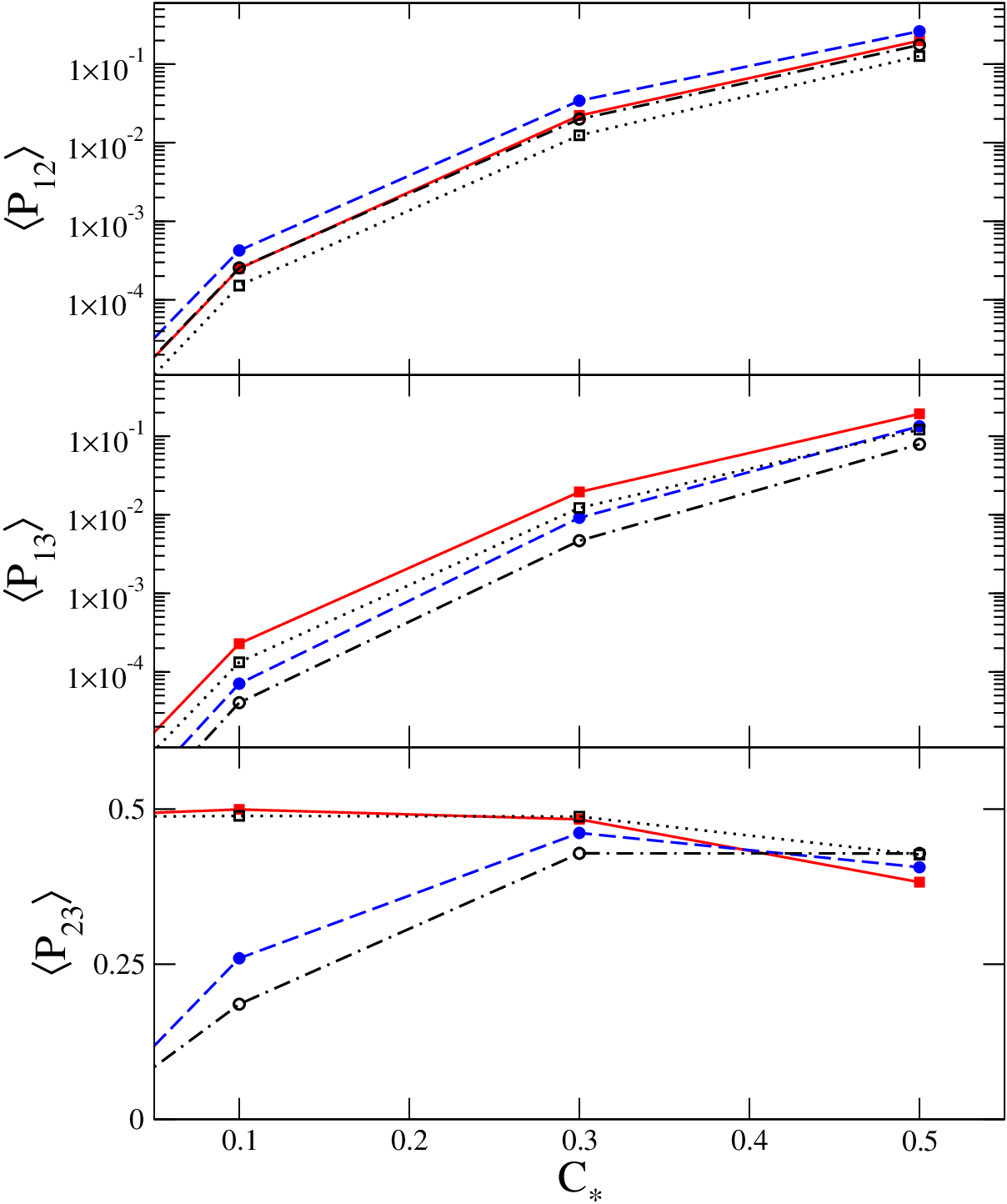}
\includegraphics[clip,width=0.45\linewidth]{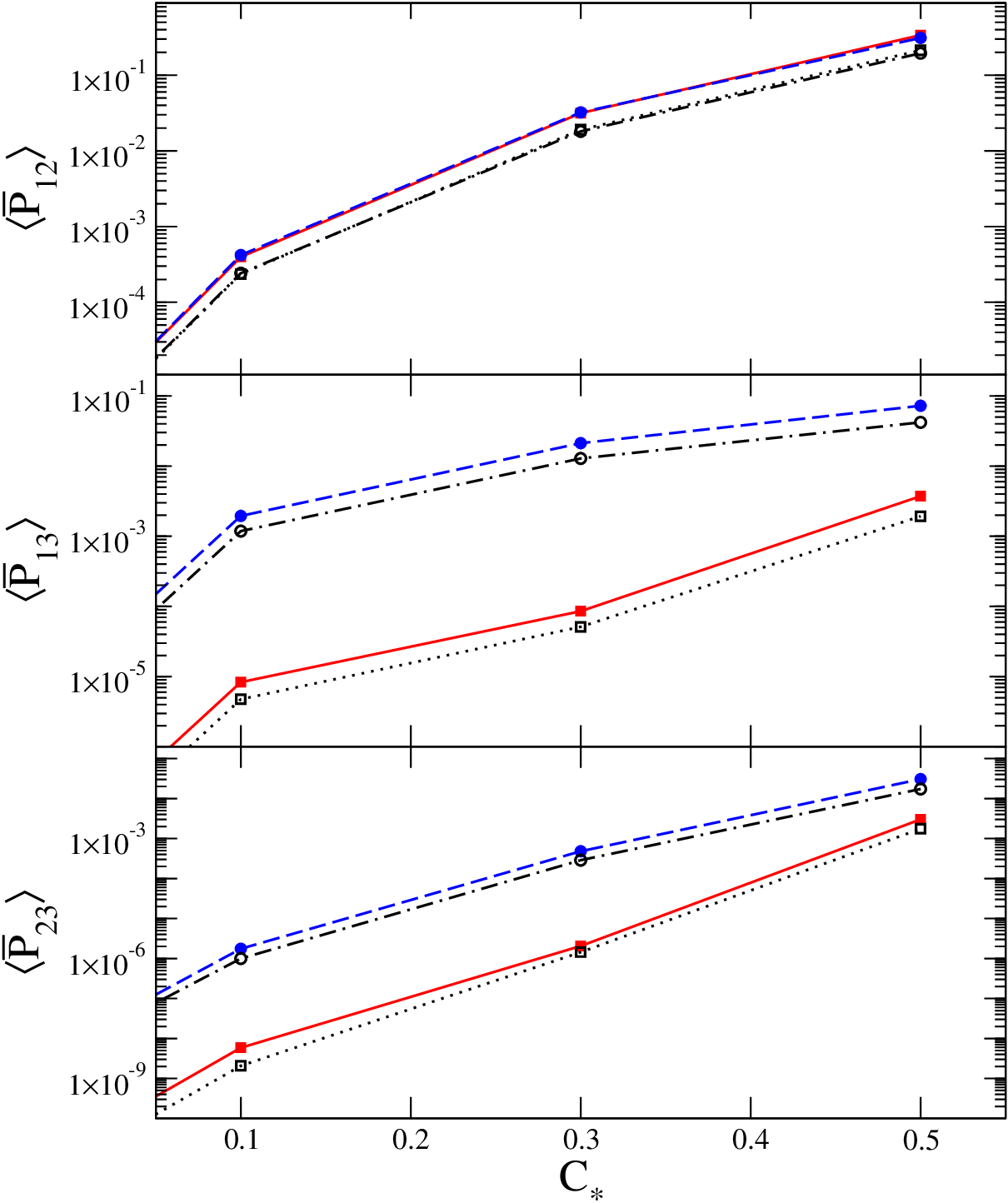}
\caption{Left figure, the mean of the transition probabilities $P_{12}$ - top panel - $P_{13}$ - center panel - and $P_{23}$ - bottom panel - 
as a function of $C_{\star}$ for neutrinos emitted from a single point. 
The right figure is the mean of the distributions for the antineutrino transition probabilities $\bar{P}_{12}$ - top panel - $\bar{P}_{13}$ - center panel - and
$\bar{P}_{23}$ - bottom panel as a function of $C_{\star}$ for antineutrinos emitted from a single point. In all panels the curves correspond to either $\sin^2 2\theta_{13}=4\times 10^{-4}$ (squares) or $\sin^2 2\theta_{13}=0.1$ (circles). The solid symbols denote our use of an anisotropic power spectrum, the open symbols to an isotropic power spectrum. \label{fig:meansvsCstar} }
\end{figure*}
We now allow the values of $C_{\star}$ and $\theta_{13}$ to float and consider both the isotropic and anisotropic power spectrum. The evolution of the 
transition probability means as a function of $C_{\star}$ for the two power spectra and two choices of $\theta_{13}$ are shown in figure (\ref{fig:meansvsCstar}).
There are many interesting trends discussed in detail in Kneller \& Mauney \cite{2013arXiv1302.3825K}. 
Large amplitude turbulence works its way through to affect every mixing channel, not just the H resonance channel $P_{23}$, as promised so that by $C_{\star} = 0.5$
we observe $\langle P_{12} \rangle \sim 20\%$, $\langle P_{13} \rangle \sim 10\%$, $\langle P_{23} \rangle \sim 50\%$, and $\langle \bar{P}_{12} \rangle \sim 20\%$, 
$\langle \bar{P}_{13} \rangle \sim 5\%$, $\langle \bar{P}_{23} \rangle \sim 1\%$.  
To put this in context, in the absence of turbulence all these transition probabilities are zero when $\theta_{13} = 9^{\circ}$.   
The only \emph{neutrino} mixing channel with reasonable sensitivity to $\theta_{13}$ is 
the H resonance channel $P_{23}$ and even then the disparity in $\langle P_{23} \rangle$ at $C_{\star} \sim 0.1$ disappears by $C_{\star} \sim 0.3$.
In contrast the \emph{antineutrinos} are very sensitive to $\theta_{13}$ even at large turbulence amplitudes: the expectation value for $P_{13}$ varies by a factor of $\sim 2$ when $\theta_{13}$ is changed from  
$\sin^2 2\theta_{13}=4\times 10^{-4}$ to $\sin^2 2\theta_{13}=0.1$, $\bar{P}_{13}$ and $\bar{P}_{23}$ on the other hand change by $\sim 1-2$ orders of magnitude between 
the same limits.  

While these trends are interesting, the purpose of figure (\ref{fig:meansvsCstar}) is to compare the use of the isotropic and anisotropic power spectra.  
Except for the H resonance mixing channel $P_{23}$, the isotropic power spectrum gives values of $\langle P_{ij}\rangle$ which are smaller than the anisotropic 
spectrum. The neutrinos are more sensitive to the turbulence when the power spectrum is anisotropic because the neutrinos are sensitive to the amplitude of the turbulence modes 
of order the neutrino oscillation wavelength \cite{Friedland:2006ta,2012arXiv1202.0776K} which is typically in the range of $\sim 10 \;{\rm km}$ in the H resonance region.   
The anisotropic spectrum removed all power for the fluctuations in the radial direction at the long wavelengths above $1/k_{\star}$ - which is of order $10^{4}\;{\rm km}$ in our calculation - and to compensate 
we needed to increase the power on the smaller wavelengths which means and effective increase of their amplitude. 
In fact we already know the exact amount the amplitude is effectively increased because we pointed out the $1/\alpha$ factor that appears in the one dimensional power spectrum in the isotropic case compared to the
one-dimensional spectrum derived from the anisotropic turbulence. 
Thankfully, our expectations are confirmed by figure (\ref{fig:meansvsCstar}) because the increase of all the mixing channels except $P_{23}$ is on the expected scale of $\alpha$.
The isotropy scale paramater $k_{I}$, which sets the scale in the radial direction below which the turbulence is isotropic, does not play a role for these point source statistics. 
The one-dimensional power spectrum along the radial direction is independent of the isotropy scale $k_{I}$ which can be seen when 
comparing equations (\ref{eq:1Diostropicps}) and (\ref{eq:1Daniostropicps}). So if the one-dimensional 
power spectrum is independent of $k_{I}$ then the effect of switching the power spectrum from isotropic to anisotropic is solely due to the removal of
radial long-wavelength fluctuations. 
\begin{figure*}
\includegraphics[clip,width=\linewidth]{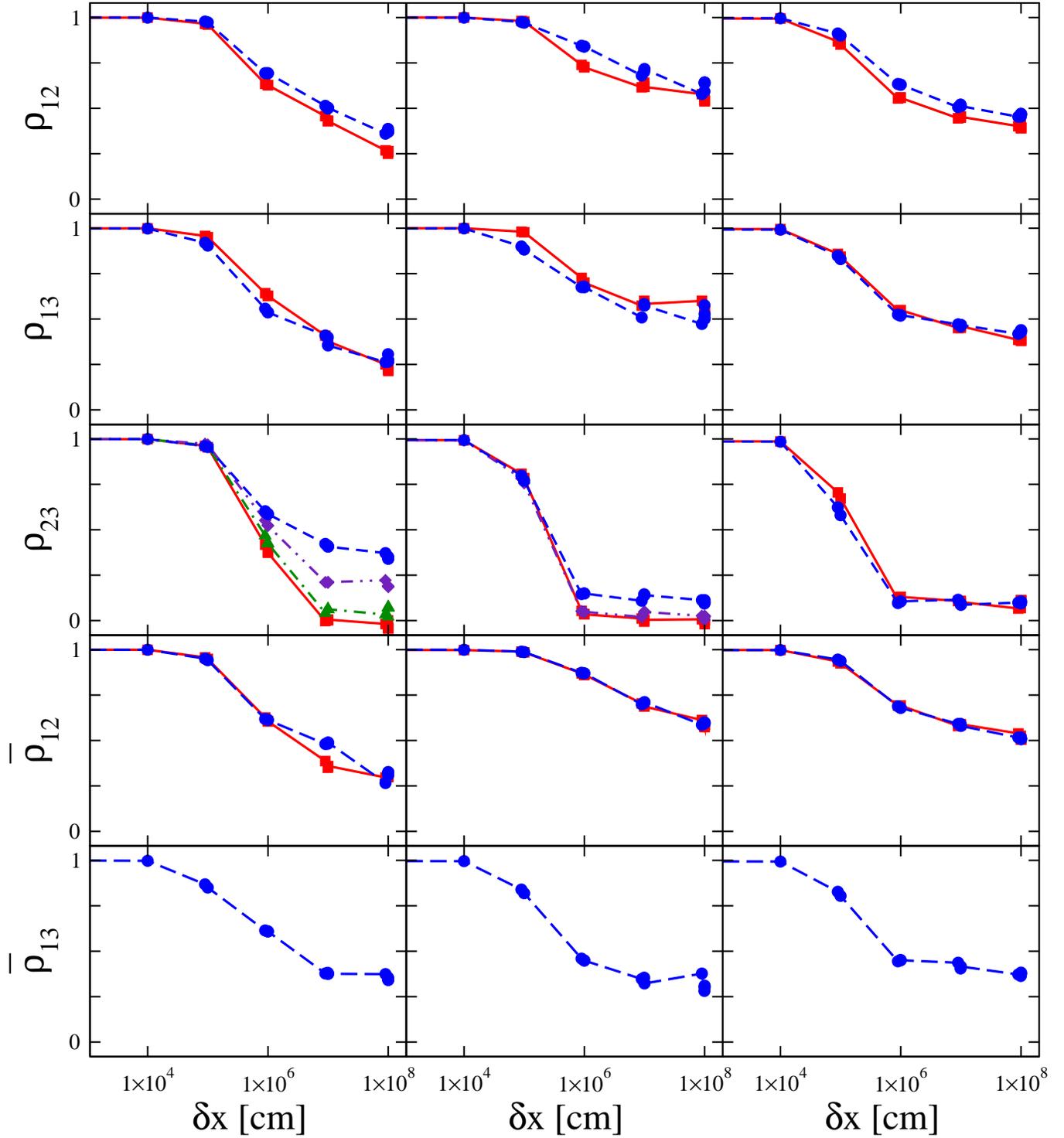}
\caption{The correlation of the transition probabilities through isotropic turbulence of
various turbulence amplitudes as a function of the distance between emission points $\delta x$. 
From top row to bottom the correlations are for $P_{12}$, $P_{13}$, $P_{23}$, $\bar{P}_{12}$ and $\bar{P}_{13}$.
The turbulence amplitudes are $C_{\star} = 10\%$ (left
column), $C_{\star} = 30\%$ (center column) and $C_{\star} = 50\%$ (right column). The values of $\theta_{13}$ are $\sin^2 2\theta_{13}=4\times 10^{-4}$ (squares joined by a solid line), $\sin^2 2\theta_{13}=10^{-3}$ (triangles joined by a dot-dashed line), $\sin^2 2\theta_{13}=4\times 10^{-3}$ (diamonds joined by a double dot-dash line) and $\sin^2 2\theta_{13}=0.1$ (circles joined by a dashed line). \label{fig:rhovsCstar}}
\end{figure*}
The transition probability $P_{23}$ behaves slightly differently but is entirely consistent with the understanding of the effects in the other channels. At
smaller amplitudes and the smaller value of $\theta_{13}$ there is no effect of the power spectrum switch upon $\langle P_{23}\rangle$ because the depolarization limit has been reached. 
At the larger mixing angle depolarization has not achieved and switching the power spectrum leads to the effects as seen in $P_{12}$ and $P_{13}$. 
The two-flavor depolarization limit is reached for the $\sin^2 2\theta_{13}=0.1$ case when $C_{\star} \sim 30\%$. At around this same turbulence amplitude there
begins the shift to three-flavor depolarization where $\langle P_{23}\rangle = 1/3$. Whatever the mixing angle used, we see that the mean value $\langle
P_{23}\rangle$ as a function of $C_{\star}$ using the anisotropic spectrum begins the transition at smaller $C_{\star}$ than the same
calculation using the isotropic spectrum because of the increased amplitude of the small scale fluctuations in the former case.

\subsection{The correlation through isotropic turbulence}
We now turn to the correlation of the transition probabilities as a function of the distance between the emission points and consider first the case of the isotropic power spectrum. 
Our result for the correlation of the transition probabilities, except $\bar{P}_{23}$, as a function of the separation $\delta x$ at various
values of $\theta_{13}$ and turbulence amplitudes $C_{\star}$ is shown in figure (\ref{fig:rhovsCstar}). $\bar{P}_{23}$ is excluded is because it 
is difficult to calculate its correlation reliably. What one notices immediately about the results are that $\rho_{12}$, $\rho_{13}$, $\bar{\rho}_{12}$ and $\bar{\rho}_{13}$ 
all show little sensitivity to either $\theta_{13}$ or $C_{\star}$  - which is in contrast to figure (\ref{fig:meansvsCstar}). The reason for the lack of sensitivity of these correlations to $\theta_{13}$ and $C_{\star}$ is explained by the exponential distributions these transition probabilities possess. Both the turbulence amplitude and the mixing angle simply `rescale' the ensemble of transition probabilities and, as equation (\ref{eq:rho}) shows, this rescaling cannot alter the correlation. One also sees that the correlation of all these transition probabilities is high, $\gtrsim 0.5$, for all separations $\delta x \lesssim 100\;{\rm km}$.  

In contrast the correlation of $P_{23}$ is sensitive to both $\theta_{13}$ and $C_{\star}$.  
When $C_{\star}$ is of order $C_{\star} \sim 10\%$ the sensitivity to $\theta_{13}$ arises because the distributions of $P_{23}$ at the different mixing angle choices are very different: 
for $\sin^2 2\theta_{13}=4\times 10^{-4}$ the distribution is uniform, for $\sin^2 2\theta_{13}=0.1$ it is strongly skewed to small values of $P_{23}$. As $C_{\star}$ increases the 
sensitivity disappears because the distributions at each value of $\theta_{13}$ become similar: this is the same behavior seen in figure (\ref{fig:meansvsCstar}). 
Finally, for $C_{\star} = 10\%$ the currently preferred value of $\theta_{13}$ gives greater correlation at a given seperation than smaller values of $\theta_{13}$. 
The correlation $\rho_{23}$ is high for $\delta x \lesssim 10\;{\rm km}$, a scale of order the
proto-neutron star diameter, at for $C_{\star} = 10\%$ and decreases rapidly as $C_{\star}$ increases. For $C_{\star} \gtrsim 0.3$ the transition probability $P_{23}$ of two neutrinos emitted
from points on the proto-neutron star separated by a distance greater than $\delta x \gtrsim 1\;{\rm km}$ are essentially independent. 


\subsection{The correlation through anisotropic turbulence}
\begin{figure}[t]
\includegraphics[clip,width=\linewidth]{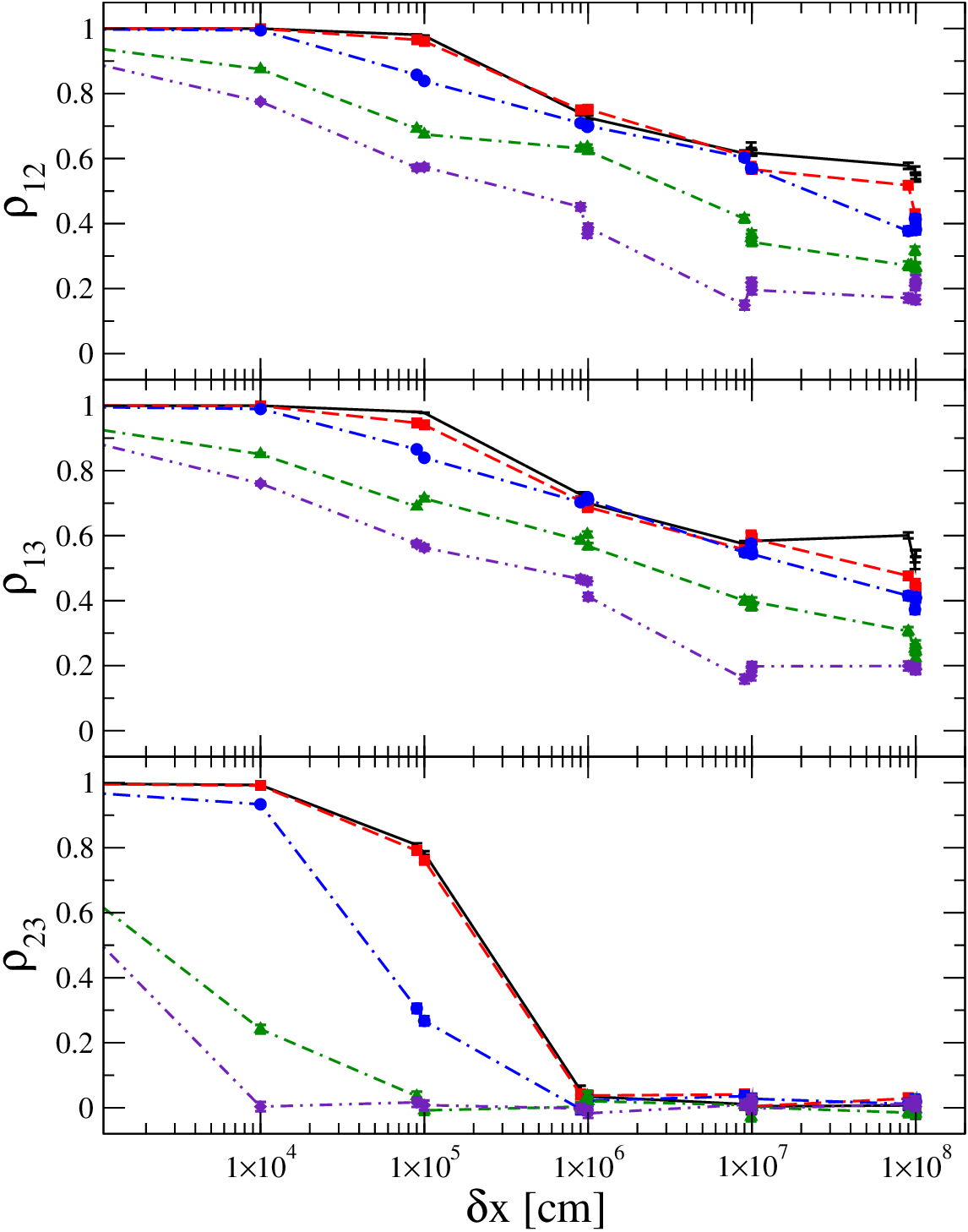}
\caption{The correlation of the transition probabilities $P_{12}$ - top panel - $P_{13}$ - center panel - and $P_{23}$ - bottom panel - through anisotropic
turbulence as a function of the separation between neutrino emission points. The turbulence amplitude is set at $C_{\star}=30\%$ and $\sin^2 2\theta_{13}=0.1$. 
In each panel the correlation of the transition probabilities through the isotropic turbulence is shown as the solid line. The other curves in each
panel correspond to different values of the ratio $f_{I} = k_{I}/k_{\star}$: $f_{I}=1$ are squares joined by long dashed lines, $f_{I}=10$ are triangles joined by
dash-dot lines, and $f_{I}=100$ are diamonds joined by dot double-dash lines. The error bars on each data point are estimated using the large N limit prediction. \label{fig:rho12,rho13,rho23aniso0.3} }
\end{figure}
\begin{figure*}
\includegraphics[clip,width=\linewidth]{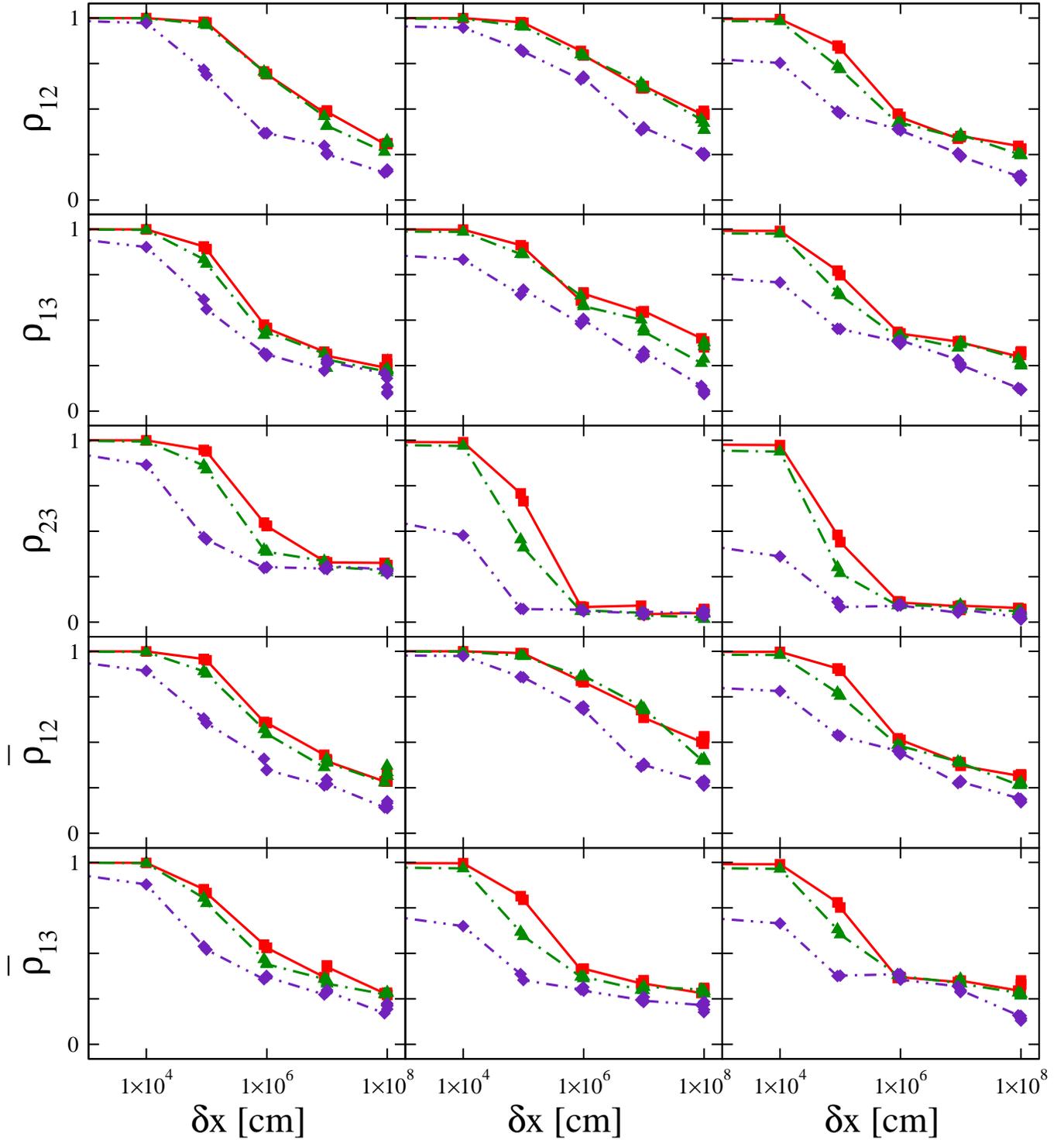}
\caption{The correlation of the transition probabilities $P_{12}$, $P_{13}$, $P_{23}$, $\bar{P}_{12}$ and $\bar{P}_{13}$ as a function of the separation between emission points $\delta x$. The
mixing angle $\theta_{13}$ was set at $\sin^2 2\theta_{13}=0.1$. The left column of panels is for a turbulence amplitude of $C_{\star}=10\%$, the central column for
$C_{\star}=30\%$, and the rightmost column is $C_{\star}=50\%$. In each panel the squares joined by the solid lines are for $f_{I}=1$, the
triangles joined by dash-dot lines are $f_{I}=10$, and the diamonds joined by dash-double dot lines are $f_{I}=100$.  \label{fig:rho12,rho13,rho23aniso9}}
\end{figure*}
The change to the mean point source transition probabilities when switching to an anisotropic power spectrum is both understandable and measurable but,
overall, the effects are small and of the order of factors of $\alpha$ i.e.~ the amplitude by which the small scale fluctuations in the anisotropic spectrum
increased in amplitude compared to the isotropic spectrum. That insensitivity no longer holds when we examine the correlations of the
transition probabilities because these quantities are functions of the isotropy scale paramater $k_{I}$. The correlations of the transition probabilities as a
function of the separation between the emission points is strongly sensitive to the amount of turbulence power in the perpendicular directions and
increasing $k_{I}$ relative to the fixed scale $k_{\star}$ shifts the power from long wavelength, small $k_{x}$ and $k_{y}$, to much shorter wavelengths, as shown in figure (\ref{fig:GRF3Daniso}). 
The effects of introducing and varying $f_{I} = k_{I}/k_{\star}$ are shown in figure (\ref{fig:rho12,rho13,rho23aniso0.3}) for the case $C_{\star} = 0.3$ and 
$\sin^2 2\theta_{13}=0.1$. For $f_{I}=1$ the difference between the isotropic and anisotropic power spectra
are minimal but at larger ratios of the two scales the correlation at some given separation $\delta x$ drops noticeably in all three
channels though the reduction in the correlation of $P_{12}$ and $P_{13}$ is not as severe as that for the transition probability $P_{23}$. It is still the case that the correlation of $P_{12}$ and
$P_{13}$ at typical proto-neutron star radii of $\delta x \sim 10\;{\rm km}$ is larger than 0.5 if $f_{I} \lesssim 10$ for this particular mixing angle choice and turbulence amplitude. 
Pushing even more power to smaller scales would lead to minimal correlation of these two transition probabilities. 
In the case of $P_{23}$, the correlation at $\delta x \sim 10\;{\rm km}$ is already small for this mixing angle and turbulence amplitude even in the
isotropic and $f_{I}=1$ cases so pushing more power of the fluctuations in perpendicular directions to smaller wavelengths completely removes the
correlation of $P_{23}$ over the proto-neutron star radial scale.
 
If we now vary the turbulence amplitude we generate figure (\ref{fig:rho12,rho13,rho23aniso9}) which is, again, for a mixing angle of $\sin^2 2\theta_{13}=0.1$. 
Examining the results we quickly observe the same general trends with changes in $f_{I}$ as seen in figure (\ref{fig:rho12,rho13,rho23aniso0.3}): increasing $f_{I}$ 
reduces the correlation with $\rho_{23}$ affected to a greater degree than $\rho_{12}$, $\rho_{13}$, $\bar{\rho}_{12}$ and $\bar{\rho}_{13}$. 
Likewise, the trends seen in figure (\ref{fig:rhovsCstar}) for changes in $C_{\star}$ are also reproduced.   


\section{Summary and Conclusions}

Supernova turbulence and its effects upon both the flavor composition of neutrinos that pass through it and their correlations depend upon many numerous parameters one needs to introduce to describe the turbulence.
All affect the result and here we try to succinctly summarize our results. 
For a neutrino energy of $25\;{\rm MeV}$ and using a supernova density profile taken from a simulation $4.5\;{\rm s}$ post-bounce, we find in a normal hierarchy that the 
correlation of the H resonance mixing channel transition probability $P_{23}$ as a function of the emission separation $\delta x$ drops considerably as $C_{\star}$ increases for both the 
cases of isotropic and anisotropic turbulence. 
If the turbulence amplitude is of order $C_{\star} \sim 0.1$ then the correlation of the transition probability $P_{23}$ for neutrinos emitted from opposite sides of the proto-neutron star, i.e.~ separated 
by $\sim 10\;{\rm km}$, is marginal for the isotropic spectrum and for the anisotropic only when $f_I \lesssim 10$.  
For $C_{\star} \gtrsim 0.3$ there is essentially no correlation of the H resonance transition probabilities for
neutrinos emitted from opposite sides of the proto-neutron star. At these amplitudes the turbulence along parallel trajectories 
separated by $\sim 10\;{\rm km}$ is just too different to permit any correlation of this transition probability in the supernova neutrino burst signal.   
If we switch to an inverted hierarchy then it will be the transition probability $\bar{P}_{13}$ which behaves this way.

In contrast, the correlation of the transition probabilities $P_{12}$, $P_{13}$, $\bar{P}_{12}$ and $\bar{P}_{13}$ in a normal hierarchy as a function of emission separation is largely independent of
$C_{\star}$ and $\theta_{13}$. The correlation decreases as the ratio $f_I = k_I/k_{\star}$ increases but remains significant for separations of order the proto-neutron star radius even for $f_I \sim 100$.  
When switching to an inverted hierarchy the mixing channels which behave this way are $P_{12}$, $P_{23}$, $\bar{P}_{12}$ and $\bar{P}_{23}$.
These mixing channels, particularly $P_{12}$ and $\bar{P}_{12}$, are the most promising for observing flavor scintillation assuming the energy resolution of our neutrino detectors does not 
wash out the effect and the temporal correlation remains high.

\acknowledgments
This work was supported by DOE grant DE-SC0006417, the Topical Collaboration in Nuclear Science ``Neutrinos and Nucleosynthesis in Hot and Dense Matter``, DOE grant number DE-SC0004786, 
and an Undergraduate Research Grant from NC State University. 


\end{document}